\documentstyle[art11,aasms4]{article}

\newcommand{\msol}{M{$_{\odot}$}} 

\newcommand{\kms}{km~s{$^{-1}$}}

\newcommand{\tco}{{$^{13}$CO}}
\newcommand{\CO}{{$^{12}$CO}}
\newcommand{\Tco}{{$^{12}$CO}}
\newcommand{\co}{C{$^{18}$O}}

\begin{document}

\title{ Multiple CO Outflows in Circinus:\\ 
	The Churning of a Molecular Cloud}

\author{ John Bally \altaffilmark{1,3}, 
         Bo Reipurth \altaffilmark{1,4}, 
         Charles J. Lada \altaffilmark{2,5},
         and Youssef Billawala \altaffilmark{1,6}}

\altaffiltext{1}{
   Department of Astrophysical and Planetary Sciences, 
   Center for Astrophysics and Space Astronomy,
   and Center for Astrobiology,
   University of Colorado, Campus Box 389, Boulder, CO 80309-0389}
\altaffiltext{2}{
   Harvard Smithsonian Center for Astrophysics, 
   60 Garden Street, 
   Cambridge, MA 02138} 
\altaffiltext{3}{ bally@casa.colorado.edu 
    | ~~http://casa.colorado.edu/$\sim$ bally} 
\altaffiltext{4}{ reipurth@casa.colorado.edu}
\altaffiltext{5}{ clada@cfa.harvard.edu}
\altaffiltext{6}{ billawal@casa.colorado.edu}

\begin{abstract}

We present a millimeter wave study of a cluster of bipolar CO 
outflows embedded in the western end of the Circinus 
molecular cloud complex, G317-4, that is traced by very high 
optical extinction.  For an assumed distance of 700 pc, 
the entire Circinus cloud
is estimated to have a mass of about $\rm 5 \times 10^4$ \msol .
The opaque western portion that was mapped in this study has a mass 
of about $\rm 10^3$ \msol , contains a number of embedded infrared
sources and various compact 1.3 mm continuum sources, 
and has a remarkable filamentary structure with numerous 
cavities which appears to be the fossil remnants of past star 
formation activity.  The most active star forming region in this
part of Circinus is centered around a compact cluster of millimeter 
continuum sources associated with IRAS 14564--6254 and IRAS 14563--6301
which lies about 7\arcmin\ to the south.  This region contains
two known Herbig-Haro objects, HH~76 and HH~77, and 
a profusion of overlapping high velocity CO outflow lobes.  Among these,
we can clearly distinguish the two largest outflows in Circinus (flows
A and B) which appear to originate from the two brightest IRAS sources. 
This region also contains at least two other prominent but 
overlapping bipolar CO outflows (flows C and C$'$), one of which may be
associated with IRAS14564--6258.
Two compact and relatively low velocity CO outflows lie at 
the northern periphery of the Circinus core and are associated with 
IRAS 14563--6250 (flow~E), a source also detected as
a 1.3 mm continuum source, and with IRAS 14562--6248 (flow G).   
A small but prominent reflection nebula associated with the nebulous
star vBH65a and a co-axial Herbig-Haro jet, HH~139, is located at the 
southeastern edge of this cloud core and illuminates part 
of a cavity seen as a low extinction region.  A faint and 
low mass CO molecular flow is associated  with vBH65a 
and HH~139 (flow~F).  The infrared source IRAS 14580--6303 drives
a small CO flow (flow~I). A second, active center of star 
formation is centered on the source IRAS 14592--6311,  the peculiar
Herbig Ae/Be star vBH65b, about 20$^{\prime}$ to the southeast of 
the main cloud core;  four HH objects, HH~140 to HH~143, and a compact 
CO outflow are located here (flow~ D).  About 5\arcmin\ further south,
IRAS 14596--6320 drives yet another outflow (flow H). 
Thus, the mapped portion of 
Circinus contains at least 10 CO emitting molecular outflows.  
Assuming that star formation has continued at a steady 
rate for the last several hundred thousand years, the Circinus cloud 
is expected to have produced dozens of young stars.
Their outflows have severely altered the structure and
kinematics of this cloud as evidenced by the multitude of 
prominent cavities and dust filaments that outline their 
boundaries.  This level of star formation activity is consistent 
with the numerous post-outflow phase H$\alpha$ emission line stars 
that have been found in this region.  The Circinus cloud complex is 
an archetypical case where star formation activity may have profoundly 
affected the structure of a molecular cloud, producing its strikingly 
filamentary and cavitated appearance and providing further evidence that
star formation may be a self regulated process.

\end{abstract}


\section{INTRODUCTION}

Star formation may be a self-regulated process in which jets, winds, and 
radiation produced by young stars may regulate the rate of gravitational 
collapse, possibly prevent collapse altogether, or even disrupt the star
formation environment.  
Recent observations have shown the ubiquity of giant, parsec-scale outflows 
(cf. Reipurth, Bally, \& Devine 1997; Devine et al. 1997; 
Eisl\"offel \& Mundt 1997; Bence, Richer, \& Padman 1996;
Bally, Devine, \& Reipurth 1996) which drive strong shocks into the 
surrounding medium where they may accelerate or even dissociate 
molecules, produce cloud turbulence, and in some cases blow out of the  
host cloud.

The Circinus cloud complex is located towards l=318$^o$ and b=$-$4$^o$ 
along a relatively extinction free line-of-sight in the southern Milky Way.  
In visual wavelength images,  the cloud stands in stark contrast 
against a rich background of stars in the fourth quadrant of the Galaxy and  
the cloud can be traced by its extinction over a region about 
2$^o$ by 6$^o$ in extent.  Our attention was drawn to this object by 
the filamentary appearance of the western portion
of the cloud which is very opaque on deep photographs (see Figure 1).
This core region contains dozens of cavities bounded 
by narrow dust filaments, about a half dozen IRAS sources, and
many signs of star formation including Herbig-Haro objects (Reipurth
\& Graham 1988; Ray \& Eisl\"offel 1994) and H$\alpha$ emission stars
(Mikami \& Ogura 1994).  A fan of nearly 
north--south oriented filaments radiates from the southern periphery 
of the main core.  A 25\arcmin\ long ridge 
of opaque material extends east--west from the southeast corner of the 
cloud core and over a dozen narrow dust lanes cross this   
ridge at nearly right angles.  The multiple cavities and filaments 
that fan away from the opaque core may indicate that 
this cloud has suffered extensive damage from the impact of 
dozens of outflows powered by young stars formed during the last 
few hundred thousand years.   

In this paper, the first of a series of multi-wavelength investigations 
of the Circinus complex are presented.   We discuss new CO 
observations that reveal the
presence of multiple high velocity bipolar outflows powered by embedded
young stars.   In subsequent papers, we will present observations of
new Herbig-Haro objects, shock excited molecular hydrogen, and a census
of the young stars produced by the Circinus cloud over the last few 
million years.

\section{OBSERVATIONS}

We observed J=$1-0$ $^{12}$CO, $^{13}$CO, C$^{18}$O
transitions during 28 April to 9 May~1988 and during 
May~1990 with the 15 meter diameter
Swedish-ESO Sub-millimeter Telescope (SEST).
We used a Schottky barrier heterodyne mixer providing a single side-band
receiver temperature of about $\rm T_{ssb}$ = 350 to 500 Kelvins
which under typical observing conditions produced system temperatures 
ranging from about $\rm T_{sys}$ = 600 to over 900 K.  
We used a chopper to compare the sky emission to an ambient
temperature absorber for calibrating the spectra and checked
the resulting calibration by comparing SEST spectra obtained at 
several locations in M17 and in the $\rho$-Ophiuchus cloud with observations
obtained with other telescopes in the northern hemisphere.
These comparisons indicate that our calibration is accurate to
about 15\%.  Observations of the compact high velocity outflow in the
Orion A core were used to estimate the main-lobe beam efficiency to be 
about $\eta = 0.7$. However, our spectra,  maps, and mass estimates 
have not been corrected for the main-lobe efficiency since most of the 
observed structures are large compared with the beam. 
The FWHM diameter of the SEST beam between 109 to 115 GHz is 
about 44\arcsec .  Spectra were recorded with a 2000 channel 
acusto-optical spectrometer that provided a channel spacing of 43 kHz.  
All observations were obtained in frequency switching mode in which 
the local oscillator was shifted by 20 MHz at a rate of about 5 Hz.
In this mode, the SEST system provided flat and stable baselines 
which for most spectra could be well fit with a first order
baseline.  Approximately 1200 \Tco , 1300 \tco, and 350 \co\ spectra 
were obtained during the two observing runs with most spectra obtained 
on uniform 40\arcsec\ grids with integration times of 300 seconds per 
position, yielding spectra with a typical RMS noise of about 0.25 K in 43 
kHz channels.  The data were reduced with the Bell Laboratories COMB data 
reduction package which was used to fold frequency switched spectra,
fit baselines, and generate contour maps, images, 
data cubes, and mass estimates of the various cloud components.

\section{RESULTS}

\subsection{Cloud Structure and Properties}

Figure 1 shows an R-band ESO Schmidt photograph of the Circinus 
molecular cloud covering about a 1 square degree field of view.  
The cloud is seen in silhouette against a rich background of stars
and consists of a complex network of opaque filaments and translucent 
cavities.  Figure 2a is a finder chart showing the locations of the
various IRAS sources and outflows discussed in the text.  The border 
of this chart is identical to the borders of the contour diagrams shown 
in Figures 2b -- d so that the locations of sources and flows can 
be easily identified on both contour maps and on the optical image
by comparison with Figure 2a.  The axes in these figures are labeled 
in arc minute offsets from a reference position 
([0,0]) which corresponds to  
$\rm \alpha (1950) = 14^h 56^m 24.9^s, ~
\delta (1950) = -62^o 54^{\prime} 59^{\prime \prime}$.

Neckel \& Klare (1980) investigated the extinction towards this line
of sight, finding an abrupt increase of A$_V$ = 0.6 magnitude at a
distance of about 170 pc and a larger jump with A$_V$ greater
than 2.0 magnitudes towards stars with distances estimated to
lie between 600 and 900 pc. 
The wall at 170 pc is unlikely to be sufficiently opaque to
be associated with the Circinus complex.  On the other hand,
the low density of foreground stars towards the cores
makes it likely that the cloud is closer than 900 pc.
We will adopt distance of 700 pc for use in estimating parameters for
the Circinus cloud and its outflows while recognizing that this value 
is uncertain by nearly a factor of 1.5.

CO emission from the Circinus cloud complex was first identified in
the survey of the southern Milky Way by Dame et al. (1987).  A 2$^o$
by 6$^o$ degree diameter CO cloud centered at velocity
$v_{lsr}$ = $-$6 \kms\ coincides with the region of optical
obscuration.  Using the integrated emission in the Dame et al. (1987) 
survey and a conversion factor of $\rm N(H_2) = 2.6 \times 10^{20} I(CO)$
yields an estimated mass of about
$\rm 4.7 \times 10^4 d_{700}^2$~M$_{\odot}$ for the mass of this
entire complex where $d_{700}$ is the distance in units of 700 pc.

Figure 2b and 2c show velocity integrated \tco\ maps showing emission
between v$_{lsr}$ = -8 to -4 \kms\ and v$_{lsr}$ = -10 to -8 \kms\ 
respectively, superimposed on the optical image of the Circinus cloud. 
Figure 2d shows a velocity integrated \co\ map showing emission
between v$_{lsr}$ = -10 to -4 \kms\ superimposed on the 
optical image of the Circinus cloud.  
Figure 3 shows a closeup of the main core 
in \tco\ integrated from v$_{lsr}$ = -8 to -4 \kms .   
The rare CO isotopes should approximately trace the
column density of molecular gas and the contour maps shown in 
Figure 2b and 2d closely match the region of high optical obscuration.

The \Tco\ data contains additional faint and narrow-line cloud components
at $v_{lsr}$ = $-$24.0, $-$9.5, $-$6.5, 
and +3.5 km~s$^{-1}$ towards Circinus.  The main components associated
with the region of obscuration shown in Figure~1 corresponds to the
bright line centered at $v_{lsr}$ = $-$6.5 km~s$^{-1}$.
None of the other \Tco\ features exhibit spatial correlations with
the region of high optical obscuration and are therefore likely to
trace low opacity foreground (3.5 km~s$^{-1}$) and 
background ( $v_{lsr}$ = $-$24.0, $-$9.5 km~s$^{-1}$) cloud
components.

Table 1 lists estimates of the dimensions, masses, and densities of
the various major quiescent components of the Circinus cloud complex.
The properties of the outflows will be discussed in the next section.  
We use the SEST \tco\ measurements to estimate the 
total mass contained within the mapped region as well as
the masses of various sub-components.  Throughout this analysis,
we assume a constant excitation temperature of T$_{ex}$ = 10 K,
which is consistent with the peak brightness of the \Tco\ line
in this region and N(H$_2$)/N(\tco ) = $\rm 7 \times 10^5$. 
We use the column density for each region to estimate the
average surface density of H$_2$ over the area of
integration (the mass surface density and number density
estimates include the correction for the cosmic abundance of helium; 
$\mu$ = 1.36).  On large scales, the various cloud sub-structures
are elongated in our maps.  We assume that the typical line-of-sight
spatial extent of each region is similar to the projected width
of each structure along its minimum dimension.   We divide this length
into the average surface density of each structure to derive an
estimate for the average number density of H$_2$ in each sub-region.  

This method probably underestimates the density and mass since 
the \co\ to \tco \ ratios in the cloud cores range from 1:3 to 1:5
indicating that along some lines of sight, the \tco\ lines may be 
optically thick.  
On average, the spectra obtained towards each local cloud maximum 
imply about a factor of 2 to 5 higher column 
density than the average column density derived in Table 1. 
Furthermore, the excitation temperature in the core regions could 
be substantially different from our assumption of 10 K.  Local heating
by embedded sources may raise the value of T$_{ex}$, which would have 
the effect of increasing the column (and volume) density derived from
our data.   On the other hand, in the interior of the cloud,
well away from external UV or local heating sources, where the 
bulk of the \tco\ emission is expected to be 
produced, T$_{ex}$ may be lower
than either in the \Tco\ photosphere (where $\tau$(\Tco) $\approx$ 1), 
or near embedded stars.  Since most of the mass in the cloud is located 
in the diffuse regions relatively far from local heat sources, these 
effects are not likely to substantially influence our mass and
average density estimates.

\subsection{Multiple CO Outflows}

The portion of the Circinus complex shown in Figures 1, 2, and 3 
contains ten molecular outflows that can be discerned in the \Tco\ 
data.  The most prominent flows lie in the main cloud core centered 
near the [0,0] position in Figures 2 and 3.  Figures 4a and 4b show 
close-ups of the main core in the \CO\ line wings which trace the 
lobes of high velocity gas emerging from this region.  
The complex morphology of 
overlapping redshifted (dotted lines) and blueshifted (solid lines) 
emission can be resolved into three or four bipolar outflows 
with opposing lobes located symmetrically about two IRAS infrared 
sources embedded within the main cloud core. 
We discuss these prominent flows in detail.  

{\it Flow A:} 
The brightest 100 $\mu$m 
infrared source in the Circinus cloud, IRAS 14564--6254 at 
the northern end of the CO core,  lies along the axis and in 
between the lobes of a 10\arcmin\ long CO flow with an axis at 
position angle (PA) of 80$^o$ measured from the source to the
centroid of the blueshifted lobe.  This flow is highly collimated 
with a major to minor axis ratio larger than 5 to 1.  
The blueshifted lobe is larger and better defined than the 
redshifted lobe.   
IRAS 14564--6254 coincides with a compact cluster
of 1.3 millimeter wavelength continuum sources with a peak
flux density of 510 mJy and an area integrated flux
density of about 2.5 Jy (Reipurth, Nyman, \& Chini 1996). 

{\it Flow B:} 
A major north--south molecular outflow is centered on a CO 
sub-condensation located about 6.5\arcmin\ further south 
near the position of IRAS 14563--6301. Flow B is highly 
collimated along a north--south axis with the blueshifted lobe 
oriented towards PA = 177$^o$.  This flow has a major to 
minor axis ratio greater than 8 to 1 and a total length of 
at least 12\arcmin .   The redshifted lobe, centered near 
[-1.2,-3] is considerably stronger than the blueshifted lobe.
Towards its northern end, the redshifted lobe of flow B is
confused with the redshifted lobe of flow A.

{\it Flows C and C':} 
In addition to the above two rather prominent outflows,
additional high velocity features are centered close to
and slightly north of the point of symmetry of flow B.
These flow components are sufficiently confused with
flow B and with each other that it is difficult to determine 
their sizes and the locations of their sources.  
The complexity of the emission indicates that at least
two highly confused flows (C and C$'$) exist in this region.
 
The flow we designate as C originates from the same region 
as flow B with a blueshifted lobe extending towards 
PA = 120$^o$.  Thus, the source IRAS 14563--6301 
may be a binary, with one component driving flow B 
and the other driving flow C.  The CO emission from flow C
consists of a group of blueshifted clumps that extend 
towards the southeast.  The redshifted counter flow is highly
confused with the red lobe of flow B, but several knots 
of redshifted emission lying to the northwest of 
IRAS 14563--6301 and to the west of the well defined B flow 
are also assumed to be associated with flow C.  
A low velocity redshifted protrusion from 
the vicinity of IRAS 14563--6301 with a major axis close to 
PA = 120$^o$ is visible in Figure 4b and may also be associated
with this flow.  Mass estimates for the redshifted lobe of 
flow C given in Table 2 are broken into two distinct spatial 
regions to avoid the contamination of the much stronger B flow. 

Figure 4a and 4b also show several redshifted and blueshifted 
knots to the north of flow C at PA = 135$^o$ that are blended with
parts of the A, B, and C flows.  The emission along this axis
switches from redshifted to blueshifted near relative offsets 
of [-1,-3.5] which corresponds to
$\alpha (1950)$ = 14$^h$ 56$^m$ 16$^s$.1 ~ and
$\delta (1950)$ = $-$62$^o$ 58\arcmin\ 29\arcsec . 
IRAS 14564--6258, which is detected at short IRAS wavelengths,
lies near this location.
Since this possible flow is highly confused with parts of 
flow C (and to a lesser extent with the redshifted lobe of the B 
flow), we label this candidate flow C$'$ in the figures.  
Flow C$'$ has a position angle of PA = 135$^o$.   Since the C and C$'$ 
flows are confused, the mass estimates in each velocity bin
listed in Tables 3 refer to the combined lobes of the C and C$'$ flows.
However, in Table 2 and 5, we have attempted to separate the total 
masses in the lobes of these flows but the resulting estimates
are highly uncertain due to source confusion. 

The proposed four bipolar outflows (flows A through C$'$) 
can account for most of the high velocity emission from the main 
Circinus core.  Each of these proposed outflows contains a 
redshifted and blueshifted lobe along a well defined outflow axis.  
For the A, B, and C flows, the point of symmetry where the 
velocity shifts from the approaching to the receding lobe coincides 
with a bright and cool IRAS source 
with fluxes increasing towards longer wavelengths that implies the 
presence of at least one class I (or earlier) embedded young stellar 
object associated with each IRAS 
source.   These sources are discussed in more detail in Section 4.
Additional outflow lobes may be hidden underneath these 
three or four dominant outflows.  

{\it Flow D:}   
The next flow to be recognized in the data
is a small outflow associated with the source IRAS 14592--6311 embedded
in the east-to-west ridge of molecular gas extending to the southeast
of the main Circinus core. IRAS 14592--6311
and flow D are associated with the peculiar Herbig Ae/Be star vBH65b.  
Four Herbig-Haro objects, HH~140 to HH~143 lie within a few
arcminutes of this region (Ray \& Eisl\"offel 1994).  

{\it Flow E:} 
Reipurth et al. (1993) reported 1.3 mm continuum emission from
two other sources in the main Circinus core.  The class~I
IRAS 14563--6250 which is located near HH~76 at offset 
coordinates [-0.7,+4.9] has a 1.3 mm flux density of 153 mJy.
A compact redshifted lobe of emission lies to the northeast and a
tongue of low velocity blueshifted emission that extends to the
southeast and blends with the outflow lobes of the A flow is 
associated with this region.  This pair of lobes are located 
symmetrically about the IRAS source and form a compact CO outflow at 
position angle PA = 220$^o$.

{\it Flow F:} 
The star vBH65a is associated with a compact reflection nebula
and a small optical jet, HH~139, that extends along the axis of symmetry
of the reflection nebula at position angle PA = 100$^o$ towards the east.
This star is associated with a class~I source
IRAS~14568--6304 and a 1.3 mm continuum source with a flux
density of 66 mJy.  Though neither Figure 4a nor 4b
shows a prominent outflow here, very low velocity CO lobes
are visible between v$_{lsr}$ = -11 to -8 km~s$^{-1}$ on the
blueshifted side of the cloud core and 
v$_{lsr}$ = -5 to -3 km~s$^{-1}$ on the redshifted side of the cloud core.
These compact lobes show elongation along the axis of symmetry of
the reflection nebula, with the blueshifted gas lying towards the
east and redshifted gas lying towards the west.  

{\it Flow G:} 
The source IRAS 14562--6246 lies about 2\arcmin\ north of 
IRAS 14563--6250 and is associated with an outflow very similar in 
size, appearance, and orientation to flow E.  This outflow
has a position angle of PA = 210$^o$.  The E and G flows are hard 
to separate from the relatively broad core of the \Tco\ spectra in 
the data cube and as a result, their spatial extents are very difficult 
to determine.  Their highest velocity emission protrudes only a few
\kms\ beyond the edge of the general \Tco\ emission of the
main Circinus core.  Thus, the mass estimates given in Table
2 and 5 are highly uncertain and may be in error by several 
factors of two.  Furthermore, the correction for hidden mass,
total momentum, and kinetic energy listed in Table 5 are
even more uncertain and should be treated as order
of magnitude estimates only.  To determine a best estimate for
the amount of mass hidden behind the cloud core, we assume that the
missing mass correction scales like that for the prominent lobes
of the A, B and C flows (see below).  

{\it Flow H:} 
At the southeastern corner of the mapped field, we found
subtle high velocity wings associated with IRAS~14596--6320. 

{\it Flow I:} 
Finally, between vBH65b and the main Circinus core lies a pair 
of infrared sources, IRAS 14580--6303 and IRAS 14582--6305.  
Subtle high velocity lobes are associated with the first and 
brighter IR source.  A small but 
prominent cavity can be seen in Figure 1 that extends due west of 
IRAS 14580--6303.  However, the orientation of the CO flow is hard 
to determine due to the small extent of our \Tco\ map near this source 
and the presence of high velocity gas throughout the mapped field.  
Using the orientation of the maximum velocity gradient in the line wings, 
we estimate that the orientation of the blueshifted lobe is about 
PA = 300$^o$, but this figure is uncertain by at least 30$^o$.

Figure 5 shows spatial-velocity cuts at several position angles through
the complex of outflows.  Figure 5a is centered on IRAS 14564--6254 
and it illustrates the kinematics of flow A.  This figure displays 
that the full width at 0.25~K is about 20~km~s$^{-1}$.
Figure 5b shows a north--south cut though flow B, which has a 
full width at 0.25~K of about 24~km~s$^{-1}$.
The A and B flows are confused on the right side of this diagram.
Figure 5c shows a cut along the axis of flow C and Figure 5d 
a cut along the proposed axis of flow C$'$.  Finally, Figure 5e shows a 
cut through the \tco\ data cube that illustrates
the typical line width of the Circinus cloud in this isotope. 

Figure 6 shows spectra of all observed CO isotopes at the 
positions of the sources of the outflows discussed above.

Our data reveal \Tco\ emission line components at several radial 
velocities.  These features limit our ability to analyze the structure of
bipolar CO outflows emerging from the Circinus cloud to velocity ranges
free from contamination by the background or foreground features.
Table 2 lists the properties derived from the analysis of the 
CO line wings from the four outflows over the velocity ranges where
contamination from gas unrelated to the Circinus cloud is minimal.
For the most prominent outflow lobes, Table 3 breaks this 
analysis into $\Delta v$ = 1 \kms\ wide radial velocity channels.

\subsection{Mass, Momentum Flux, and Kinetic Energy Estimation}

We use two procedures to estimate the outflow masses, momenta,
and energies, one well known, and the other new.  
The first procedure simply assumes that I(\Tco )/I(\tco ) 
is {\it independent} of the radial velocity of the gas
and has a mean value determined by integrating the
\tco\ emission over the spatial and velocity extent of the
\Tco\ emission from each lobe of an outflow in a velocity range
that excludes the emission from the cloud core.
In contrast, the second procedure fits a polynomial function to the
{\it velocity dependence} of the I(\Tco )/I(\tco ) ratio in the average
spectrum of each outflow lobe in the portion of the line profile
that traces the high velocity emission. 
This function is used to extrapolate the I(\Tco )/I(\tco ) ratio 
as a function of velocity into the far line wings where the 
observed intensity of \tco\ is too faint to
be directly measured.  Though these procedures produce 
similar total mass estimates, the resulting {\it velocity dependences}
of the mass, momentum, and flow kinetic energy are
different.

\subsubsection{Assume Isotope Ratio Is Independent of Velocity}

The signal-to-noise ratio in the high velocity outflow lobes
is much higher in the \Tco\ data than in the \tco\ data.
Therefore, the \Tco\
spectra are used to estimate the mass in each outflow lobe and
in each velocity bin.  The area averaged and velocity integrated 
\tco\ and \Tco\ intensities are measured over the spatial extents 
of each outflow lobe as determined from their extents in the \Tco\ 
maps.  In the line wings (excluding the emission from the cloud
core) the area and velocity averaged ratio, 
$\int _{area} \int _{velocity}$ I(\Tco )/I(\tco ), is found 
to vary from about 12 to over 33 with a median value of 20, 
indicating that the outflow lobes are optically thick in \Tco\ 
with a mean optical depth of about 4. We use this constraint 
to estimate the mean opacity of the \Tco\ line in each lobe
and to correct the \Tco\ based mass estimates for optical depth.  
In practise, this is achieved by dividing the \Tco\ integrated fluxes
(integrated over the area of each outflow lobe and the velocity 
interval of interest) by 20, the mean I(\Tco ) / I(\tco ) ratio,
and treating the resulting spectrum as if it were a \tco\ spectrum
(equivalent to estimating the \tco\ spectrum from 
the \Tco\ data).  The mass in the high velocity lobe is 
computed from the scaled (by the mean \tco / \Tco\ ratio) \Tco\ flux 
integrated 
over the solid angle and velocity range of the outflow lobe 
using the same assumptions as for the analysis of the \tco\ data 
for the quiescent cloud.
Estimates of the mass, momentum, and kinetic energy of each major
outflow lobe using this method are presented in Tables 2 and 3.  

For the most prominent outflow lobes, the analysis is done over a set
of $\Delta$v = 1 \kms\ wide radial velocity intervals from the highest
velocity (with respect to the cloud core) where the line wings 
can be seen to the lowest velocity at which the line wings are free 
from contamination from the cloud core or from unrelated cloud components.
The area and velocity integrated \Tco\ emission is scaled in the COMB
data reduction package to produce the mass in each velocity and area 
interval.   These are listed in Table 3.  

We assume that the flows are each inclined by an angle of $i$ = 60$^o$ 
with respect to the line-of-sight (the statistically most probable angle 
for random flow directions) so that the observed radial velocity is
exactly one half of the true velocity.  Thus, as an example, 
the mass observed in the
radial velocity interval v$_{lsr}$ = $-$16 to $-$15 \kms\ is 
displaced from the cloud core velocity (v$_{lsr}$ = $-$6.5 \kms ) by an 
observed radial velocity difference of $\rm \Delta v_{rad} = 9$ \kms .  
When corrected for the assumed inclination angle,
this mass element has a physical velocity of 
$\rm \Delta v = \Delta v_{rad} / cos(i) = 18$ \kms .  The momentum and
kinetic energy associated with the motion of the mass in a given 
radial velocity range with a mean velocity displacement $\Delta v$
is given by P = M($\Delta v$)~$\Delta v$ 
and E = $\rm 0.5 M(\Delta v) \Delta v^2$. These inclination corrected 
velocity differences are used to calculate P and E in Tables 2 and 3 
for each calculated M($\Delta v$),  the mass observed
in the specified velocity interval integrated over the solid angle 
indicated in Table 3.  The total {\it observed} lobe momenta and kinetic 
energies listed in Table 2 are obtained by summing over 
the observed velocity ranges and are {\it not} corrected for the 
contribution of the velocity ranges over which the outflow 
lobes are hidden by the foreground and background emission from the cloud. 

The mass, momentum, and kinetic energy in each $\Delta v$ = 2 \kms\ wide 
velocity interval (corrected for inclination) can be represented as 
a power law dependence on the flow velocity with the form 
$M(\Delta v) = M_0 \Delta v ^{- \alpha }$,
$P(\Delta v) = P_0 \Delta v ^{- \beta }$, and
$E(\Delta v) = E_0 \Delta v ^{- \gamma }$ over the velocity range of the
measurements.  
The redshifted and blueshifted lobes have slightly different power law 
indices. The mean power law indices (averaged over the five best  
defined lobes) are: 
$\alpha = 2.7 \pm 0.2$,
$\beta  = 1.7 \pm 0.2$, and
$\gamma = 0.7 \pm 0.2$.  

We use our power law relations derived above
to estimate the amount of mass, momentum,
and kinetic energy hidden by the bright emission associated with the
quiescent cloud.  The masses, momenta, and kinetic energies in Tables 
2 and 3 were estimated over a range of inclination-corrected radial 
velocities extending from close to the redshifted and blueshifted
edges of the \tco\ and \co\ lines where the particular outflow lobes
become difficult to distinguish from the wings of the \Tco\ line
($\Delta v_2$; this and the other $\Delta v$ parameters refer to the 
{\it difference} between the inclination corrected velocities of 
a fluid element and the line centroid velocity of the cloud core) to 
the velocity where the \Tco\ emission is lost in the noise 
($\Delta v_3$).  In the A, B, and C~+~C$'$ flow lobes,   
$\Delta v_2$ = 6 \kms\ on the redshifted side and 
$\Delta v_2$ = 8 \kms\ on the blueshifted side of the core and 
$\Delta v_3$ = 18 \kms .  We seek to estimate the amount of mass, 
momentum, and kinetic energy that lies hidden between $\Delta v_2$ 
and the lowest velocities where a parcel of gas is likely to be still 
considered as part of the outflow, ($\Delta v_1$) set by the 
turbulent velocity of the core.  We estimate $\Delta v_1$ from the 
full width half maximum (FWHM) width of the \tco\ and the width at the 
base of the \co\ lines, finding that in the A, B, and C+C$'$
lobes, $\Delta v_1$ = 2.0 \kms (see Figure~6).

The ratio of total mass in the interval $\Delta v_1$ to $\Delta v_3$ 
to the observed mass in the interval $\Delta v_2$ to $\Delta v_3$
is given by 

$$ { M_{total} \over M_{obs}} = 
   {
     {(\Delta v_3 / \Delta v_1)^{\alpha -1} - 1 }
	    \over
     {(\Delta v_3 / \Delta v_2)^{\alpha -1} - 1}
   }
$$  

\noindent
For the above values of $\Delta v_1$, $\Delta v_2$, and $\Delta v_3$ 
and a mass power law index $\alpha$ = 2.7, the best estimate of the 
true mass of gas in each outflow lobe is about 10 times greater than 
the observed mass ($M_{obs}$) in the velocity range between 
$\Delta v_2$ and $\Delta v_3$ listed in Table 2.  
The correction factor for the momentum is about 5 and for kinetic
energy is about 2.2.  Applying these corrections for hidden mass to the
lobe parameters in the top of Table 2 leads to the total mass, 
momenta, and kinetic energy estimates listed in Table 5.

\subsubsection{Polynomial Extrapolation of the Velocity Dependence 
of the Isotope Ratio}

The second, new method of analysis to estimate the masses 
in each outflow lobe is based on the realization that the \tco\
observations also can trace the low velocity inner wings of the 
outflows.   Averaged over the \Tco\ determined spatial extents of the
redshifted and blueshifted lobes of the A, B, and C+C$'$ flows,
the \tco\ emission can be used to estimate lobe masses wherever the
lobe-averaged \tco\ emission line exceeds one standard deviation above
the RMS noise for the averaged spectrum of the flow in question (0.012~K
for Flow A).  Between the velocity where this criterion is met, and the velocity
where the emission from the turbulent cloud core dominates the line
profile,  we estimate the lobe column density and mass by 
assuming that \tco\ is optically thin.

At velocities where the \tco\ emission, integrated over the spatial 
extent of each outflow lobe as determined from the \Tco\ emission, 
falls below the level where it can be reliably measured 
(below 0.012 K for Flow A), the \Tco\ data must be used for column 
density and mass estimation.   However, the \Tco\ emission must
be corrected for optical depth.  Instead of assuming that the
isotope ratio is constant and independent of velocity as in
the first mass estimation method, we extrapolate the velocity dependence 
of the isotope ratio from the inner line wings where its velocity 
dependence can be reliably determined.  A second order polynomial is 
fit to the velocity dependence of the isotope ratio over the velocity 
regime where
the \tco\ line is reliably detected in each outflow lobe (see Figure 7).
Only those data that trace gas outside the line core
and with $\int _{area} W(^{13}CO)$ and $\int _{area} W(^{12}CO)$
greater than their respective RMS noises are used in the polynomial fit. 
The resulting fit, $R_{12/13}(v)$, is used to extrapolate the isotope
ratio into the velocity regime where only the \Tco\ line is reliably 
detected. However, for the purpose of extrapolation, this function is 
truncated when it reaches a value of 89, the assumed intrinsic \Tco\ to 
\tco\ isotopomer ratio. 
We thus replace the {\it observed} \tco\ line profile with the 
{\it estimated} \tco\ profile 
obtained by calculating the quantity T(\Tco ) / $R_{12/13}(v)$ 
for those velocities where \Tco\ is reliably detected, but 
the \tco\ line falls in the RMS noise criterion.   

Results of this second procedure are also tabulated in Tables
2, 3, and 4.
Power law fits to the mass as a function of velocity in each
outflow lobe are shown in Figure 8.  Only those points that 
do not lie within the cloud core velocities (-8 \kms\ to -5 \kms ) 
are used in the fit.   For each lobe, the mass functions for
both redshifted and blueshifted gas are plotted but the 
more prominent wing is used to label each outflow lobe as 
red-- or blueshifted.   

\section{Discussion}

\subsection{Multiple Outflows}

Several dozen IRAS sources located towards the Circinus complex
are listed in Table 4.  
The brightest 100 $\mu$m sources are associated with CO outflows.  
The driving source of flow A is IRAS 14564--6254, 
which was observed at 1300 
$\mu$m by Reipurth, Nyman \& Chini (1996). They clearly resolved 
the object into 4 distinct sources, suggesting that a small stellar 
aggregate is being formed there. An H$\alpha$ emission star is 
located close to the IRAS position (star 5 in Mikami \& Ogura 1994),
but this objects is almost certainly unrelated to the IRAS or
1300 $\mu$m sources. The B and C flows have axes which intersect almost 
precisely at their centers of symmetry and at that location 
we find the class I source IRAS 14563--6301. 
This suggests that the source is multiple, 
with at least two components simultaneously driving outflows. 
A chain of redshifted and blueshifted lobes that is confused with
the C flow and with the redshifted lobe of the B flow may delineate a
flow (C$'$) centered at [-1,-3.5]. At least three more much fainter flows
(E, F, and G) are associated with IRAS sources and are nearly lost behind 
the overlapping lobes of the  stronger outflows and the broad emission
from the relatively turbulent main Circinus core.  
IRAS 14563--6250 near the 
Herbig-Haro object HH~76 drives flow E and IRAS 14568--6304 
(near vBH65a and HH~139) drives a very faint flow F.  
Both IRAS sources are 
also associated with compact 1.3 mm continuum sources. 
Flow G lies  several arc minutes north of flow E and is very similar to it
in size, appearance, and orientation.  Three additional flows lie east 
and south of the main core.  
The source of flow D, IRAS 14592--6311 coincides 
with a visible star with bright reflection nebulosity, vBH 65b 
(van den Bergh \& Herbst 1975), which is associated with four HH objects 
(Ray \& Eisl\"offel 1994).  Flows I and H are driven by IRAS sources
embedded in a prominent dust condensation easily seen in Figure 1.

It is interesting to note that a prominent and visually very opaque
core near [14,-12]
($\alpha (1950)$ = 14$^h$  58$^m$    28.7$^s$,
$\delta (1950)$ = $-$63$^o$ 06\arcmin\ 59\arcsec )
does not harbor neither a known IR source nor exhibits
\Tco\ line wings indicative of an outflow.
Its high opacity, compact size,
and strong \co\ and \tco\ emission indicate that it is very dense.
The integrated line intensity is 4.6 K~km~s$^{-1}$ in \tco\
and 1.0 K~km~s$^{-1}$ in \co , a ratio characteristic of the densest cores
in Circinus and indicating that \tco\ may be optically thick.
It is possible that this condensation is still in a pre-collapse phase
and may therefore be an ideal candidate for the investigation of a
cloud core in a pre--star forming state.

\subsection{Star Formation in Circinus}

We can estimate the time scale for star formation in Circinus, 
the number of stars formed in the Circinus complex, 
and the cloud star formation efficiency from the observed outflow
and cloud parameters.  The field included in the \tco\ SEST map is
estimated to have a mass of about 900 \msol\ (see Table 1) 
and a turbulent line width of $\Delta v$ = 2.5 \kms\ estimated by 
averaging all of our \tco\ spectra.   If the product of these 
two numbers, $\rm P_{cloud} = 2.2 \times 10^3$
\msol\ \kms , is the result of acceleration of the cloud by 
the radiative shocks powered by $N$ outflows, the number of outflows 
required to generate the observed motions is $\rm N = P_{cloud}/P_{flow}$.  
The four most massive flows discussed here in detail (A, B, C, and
C$^{\prime}$) dominate the 
mass, momentum, and energy injection into the cloud (see Tables 2 and 5).
They have an average observed $P_{flow} = 30$ \msol\ \kms\ in each flow.  
This value sets a lower bound 
since it is uncorrected for hidden mass and excludes the
impact of the smaller flows.  Table 5 lists estimates for 
this quantity corrected for hidden mass and these values 
imply $P_{flow}$ = 200 \msol\ \kms .  
Thus, this range of values for $P_{flow}$ implies that between 11 to
73 stars similar to the driving sources of the three or four most massive
flows are required to have formed in the portion of Circinus
mapped with the SEST to produce the observed internal motions.
When we include lower mass stars that power some of the weaker
flows, the estimated number of stars required to have formed may range 
from 25 to over 100 stars. 
If the median stellar mass associated with the more powerful flows is
1 \msol , the implied star formation efficiency in this portion
of the Circinus cloud,  
$\eta _{SFE} = M_{cloud} / M_*$, ranges from 1.2 to 8\% when only 
the most massive stars (such as those that power the A, B, C, and C$'$
flows) are considered, or $\eta _{SFE}$ = 5 to 20\% when the sources
of all 10 outflows are considered. 
If the three or four most massive flows discussed represent a typical 
steady state, and outflows have a lifetime of $10^5$ years
(Bally \& Lada 1983), then the duration of this star formation 
activity is about
$\tau _{SF} = N_* \tau _{flow} / N_{flow} \approx $0.5 to 2 
$\times 10^6$ years
for $N_*$ = 25 to 100 stars, and $N_{flow} = 5$ active flows
present at any one time.
Since some of the outflows may have blown clear of the cloud and deposited 
their flow energy in the surrounding lower density intercloud medium, this
estimate is a lower limit.   However, if the outflows do blow out of the
parent cloud, the cavity walls left behind still absorb a substantial
fraction of the energy of the flow and, at most, blow-out will increase
the required number of stars and the star formation efficiency by
about a factor of 2.

Some of the weaker flows may be much older than the most active flows.
For example, flow F appears to be associated 
with a fairly massive star, vBH65a, which may be similar in mass
to the sources of the more prominent flows.  However, this star,
and its outflow may be much older than the A, B, C, and C$'$ flows
since the star is visible, has associated visible HH objects, 
and may have formed a large cavity devoid of molecular gas.   
Therefore, the dynamic ages of some of the weaker flows may 
under-estimate the true age of the source star or its outflow.
  
Mikami \& Ogura (1994) found 14 H$\alpha$ emission line stars 
towards the portion of the Circinus clouds covered by our survey.
Reipurth (private communication) has identified 
additional but fainter emission line stars.
These stars provide evidence for previous star formation activity and
when these sources were embedded young stars, they 
probably produced flows similar to those now observed.  
As the outflow from each source subsides, the ejected and  
entrained gas decelerates in momentum conserving interactions with the 
ambient medium to form a dense shell of swept-up gas surrounding
cavities of low density material.  The prominent dust filaments may 
correspond to the fossilized walls of previous generations of outflows, 
some of which may have been powered by the visible young stars in this 
field.

\subsection{Cavities as Fossil Outflows}

Figures 2 and 3 show that the dust filaments correspond to
prominent features visible in the \tco\ and \co\ maps. 
Thus, these structures have relatively high column densities with
N(H$_2$) ranging from 10$^{21}$ to 10$^{22}$ cm$^{-2}$.
Inspection of the radial velocity field in \tco\ shows that most of
these filaments have velocities of less than 1 \kms .  
Many of the filaments are about 1 pc long and about 0.1 to 0.2 pc wide 
and enclose cavities with dimensions of about a parsec.  
A good example of a prominent cavity lies along the axis of the 
HH~139 jet associated with vBH65a.  This cavity may be young compared 
to the other cavities evident in Figure 1, but old compared to the
other active outflows in Circinus since it is associated
with an active stellar source, a jet, but only a very weak CO outflow. 

Our \tco\ data shows a 2\arcmin\ by 5\arcmin\ clump
of emission at v$_{lsr}$ = -9 \kms\ centered about 4\arcmin\
southeast of vBH65a at [6,-12] and elongated along an axis 
pointing back towards this star.  This feature coincides 
with a nearly north--south (PA = 165$^o$) and opaque filament 
of dust at the southern periphery of the cavity that may by 
associated with vBH65a.  A second but fainter
clump with a similar size 
and elongation is located at [11,-16]. These features trace 
an anomalous velocity component that is blueshifted by about 
3 \kms\ with respect to the emission produced by the bulk 
of the \tco\ emission from the Circinus complex. 
Nowhere else is a component at this velocity observed in 
the mapped field.  One possible interpretation is that this
anomalous \tco\ emission component traces gas that has been 
expelled a long time ago by an ancient outflow.  If the source
was vBH65a, then the outflow orientation from this
source must have changed by about 25$^o$,  the dynamical age 
of the feature centered on [6,-12] is about $2 \times 10^5$ 
years, and the age of the feature at [11,16] is at least 
$5 \times 10^5$ years.

For cavity dimensions and wall velocities of 1 pc and 1 to 2 \kms\, 
the cavities are expected to survive for about 0.3 to 1.0 Myr 
before the random motions in the cloud distort their shapes or 
fill in their voids.  Thus, the cavities have lifetimes about an 
order of magnitude longer than the dynamic ages of the visible 
CO outflows.  Thus, if star formation has continued at a steady 
state for a significant fraction of a million years, there 
ought to be dozens of cavities and filaments that delineate 
cavity walls.  This is consistent with the presence of dozens of 
prominent dust filaments in the Circinus cloud which may therefore 
trace the walls of cavities produced by old and long extinct 
fossilized outflows. Their numbers are consistent with the above 
predictions for the number of young stars produced by the 
Circinus cloud and with the number of observed H$\alpha$ emission 
line stars.

\subsection{Emission Line Wing Power Law Indices }

Several investigators have found that the high velocity \Tco\ emission 
in the lobes of bipolar outflows can be characterized by a broken power law
(cf. Lada \& Fich 1996; Lada \& Fich 1997, 1998;
Shepherd et al. 1998; Bachiller 1996; Stahler 1994) with a mass index
$\alpha = 1.8$ at low velocities and $\alpha > 4$ above the break.
Zhang \& Zheng (1997) have modeled the structure of the CO
line wings in the optically thin limit for models of the molecular outflow 
in which the emitting gas is entrained from the surrounding medium by 
a bow shock.   The resulting models produce broken power laws with 
an index of $\alpha = 1.8$ at low velocities and $\alpha = 5.6$ above 
the break as observed in a variety of bipolar molecular outflows.
One interpretation of this result is that the radial velocity where
the break occurs is related to the location of the emitting gas with respect
the the bow shock which entrains the surrounding medium.  
Gas below the velocity where the break occurs is associated with the 
low velocity material entrained by the far bow shock wings while 
gas at higher velocities is associated with the mostly forward moving 
head of the bow shock.  However, in considering the application of the 
Zhang \& Zheng models, it is important to remember that {\it optically 
thin} high velocity emission is assumed.  The line wings produced
by the Circinus outflow lobes are demonstrably optically thick in the 
lower velocity portions of their lobes.

The Circinus power law indices derived by the first method of mass
estimation are higher (by about 1.0 in the slope) than the power law 
indices derived for the inner wings of many other bipolar outflows 
(Stahler 1994; Bachiller 1996), but considerably 
lower than the indices found for the highest velocity gas at velocities 
beyond the velocity where the power law breaks.  
The indices in Circinus imply that most of the mass 
of each outflow lobe lies at the lowest velocities that are likely to
be hidden by the cloud core.   

The second method of mass estimation produces power law indices that
are much steeper than the first method with indices that
range from nearly 3.0 to slightly over 5.0, about 1 to 3 times 
steeper than the indices found by using the first mass estimation method.  
This difference is a direct result of the assumptions made about 
the variation of optical depth with velocity.  In the second 
method, the polynomial fit results in rapidly decreasing line 
optical depth with increasing velocity, while in the first method, 
the optical depth is assumed to be independent of 
velocity.  

The second method produces power law indices which approach the
power law indices found for outflows at velocities above the
velocity where the power law index breaks.  This suggests that
the broken power law indices found for other outflows may be an
artifact of the assumptions used in the analysis of
the mass functions.  The low power law indices in the inner 
portions of the line profiles may trace those parts of the flow
where the \Tco\ emission is optically thick and the steeper
outer power law indices may trace the parts of the  
outflow where the \Tco\ emission is optically thin. 
The momentum and kinetic energy power law indices can be
found by adding 1.0 or 2.0 to the indices shown in Figure 8. 
Future measurements of the velocity structure of the lobes of 
outflows with the arc second angular resolution provided by
interferometry are needed to clarify the origin of the
observed power law indices and of the broken power laws seen
in some CO outflows.

\subsection{Dynamical Ages and Momentum Injection}

For the fastest visible gas ($\Delta v$ = 18 \kms ; corrected for
inclination), the dynamical ages of these flows 
($\tau$ = R / $\Delta v$) ranges from $1.4 \times 10^4$ years (flow D) to
$6.7 \times 10^4$ years (flow B).  On the other hand, the lowest observed 
velocities have dynamical ages about 2 to 3 times longer and gas that
is just moving above the turbulent velocity of the core and remains
hidden has a dynamical age about 4 to 10 times longer.   As with other
bipolar CO flows, these estimates are highly uncertain 
(to a factor of 2 to 3) and depend in detail on the 
nature of the flow dynamics that is assumed. 

The observed momentum and kinetic energy can be divided by the
dynamical ages of each lobe to estimate the momentum injection rate
and mechanical luminosity of each outflow.
Using the inclination and hidden mass corrected values in Table 5, and
an assumed flow lifetime of about $10^5$ years implies that 
each of the three major Circinus outflows entrain mass at a rate 
$\dot M \approx 1 \times 10^{-3}$~M$_{\odot}$~yr$^{-1}$,  
they inject momentum into the cloud at a rate
$\dot P \approx 3.5 \times 10^{-3}$~M$_{\odot}$~\kms\ yr$^{-1}$,  
and the resulting mass motions have a mechanical luminosity of 
$\dot E \approx 1.5$~L$_{\odot}$. 
The corrected outflow masses are comparable to the
total masses estimated for the cloud cores in which the sources
are embedded.  This implies that at the presently observed stage, 
a large fraction of the core masses may be influenced by the mass 
flows driven by the YSOs.

\subsection{The Churning of the Circinus Cloud} 

The Circinus cloud appears to have been severely modified by
extensive star formation over the past few million years.  The
formation of dozens of stars has shredded and churned the cloud, 
producing dozens of fossil outflow cavities surrounded by dense 
filaments of compressed gas  that may be responsible for the 
``Swiss cheese'' appearance of the cloud.  Even with low efficiency, 
star formation is capable of producing the observed chaotic motion and
structure in the cloud.  

We have come to recognize that star formation is 
highly correlated and that even in low mass clouds, stars form in 
very dense groups that frequently produce multiple overlapping 
outflow lobes.  When our first Circinus data were obtained nearly 
10 years ago, we did not have the fortitude or confidence to attempt 
to resolve the confusion of overlapping outflow lobes and line wings 
into individual outflow lobes.  In part, we were hindered by the lack
of efficient three dimensional visualization tools needed
to analyze the phase-space of our observed data cubes. But
more significantly, we lacked the experience that has recently been
gained from the detailed investigation of environments such as
OMC~2/3, NGC~1333, and Cepheus~A, where dozens to hundreds of young
stars produce a confusion of overlapping outflows. The arc second 
angular resolution of visual and near infrared-images was needed
to resolve the confusion in these regions and has greatly aided in
the deconvolution of overlapping outflow lobes in the millimeter 
wavelength maps 
of these regions.  These data have taught us to first identify 
the dominant flows, then to proceed to search for ever weaker  
and more subtle signs of additional but fainter flows, especially in 
the vicinity of very cool IRAS sources.    It is this procedure
that has finally enabled us to understand the structure of the
Circinus outflow complex. 

\noindent
{\bf Acknowledgements:}  We thank David Theil for assistance in
the preparation of the figures.

\section{Appendix:  Mass and Column Density Estimation Formulae}

Here, we provide the formulae which describe the procedures used 
in estimating column densities and masses from the data. 
The column density is given by (cf. Margulis \& Lada 1985)

$$ { N(^{13}CO)} =
    {
      2.5 \times 10^{14} {({T_{ex} - 0.91}) \over {T_{ex}}}
      { {\int T_A^{13}(v) dv} \over {1 - exp(-T_o/T_{ex})} }
    }
$$

\noindent
where the excitation temperature $T_{ex}$ is given by

$$  { T_{ex}} = 
    {
	5.5 \over log \left[1.0 + 
        \left( {5.5 \over {T_{max}(^{12}CO) + 0.82}}\right) \right]
    }
$$

\noindent
with $T_o$ (= $h \nu /k$) = 5.3 K for the \tco\ line. The mass is 
given by

$$  { M(H_2)} = 
    {
	2 \times \mu \times m_H \times  N(H_2) \times A
     = 2.2 \times 10^{-16} \times N(H_2) \times d^2_{100} \times 
     A_{sr}  ~~~~~ M_{\odot}
     }
$$

\noindent
where $\mu = 1.36$ is the mean molecular weight of hydrogen corrected for 
helium and other trace constituents, $m_H$ is the mass of hydrogen, 
$N(H_2)$ is the $H_2$ column density, and A is the spatial area of the
flow $A = 4.68 \times 10^{42} d^2_{100} A_{sr}$ $pc^2$,
$d_{100}$ is the distance to Circinus in units of 100 pc, and $A_{sr}$ 
is the area of the flow in steradians, and 
$N(H_2) = 7 \times 10^5 N(^{13}CO)$ $cm^{-2}$.

When only the \Tco\ data is available, we estimate masses from this
line by estimating the equivalent \tco\ line using the assumed
\Tco\ / \tco\ ratio (using either of the methods described in
the text;  this is equivalent to correcting the \Tco\ based
column density or mass estimate for optical depth).
The estimated \tco\ profile is obtained 
by calculating the quantity T(\Tco ) / $R_{12/13}(v)$ 
for those velocities where \Tco\ is reliably detected, but 
the \tco\ line falls in the RMS noise criterion.   
Thus, \tco\ column density estimated from the \Tco\ profiles
is given by 

$$ { N(^{13}CO)} =
    {
      2.5 \times 10^{14} {({T_{ex} - 0.91}) \over {T_{ex}}}
      { {\int T_A^{12}(v) dv} / {\int R_{12/13}(v) dv}
      \over {1 - exp(-T_o/T_{ex})} }
    }
$$
.

\section*{REFERENCES}

\noindent 
Adams, F. C., Lada, C. J., \& Shu, F. H. 1987, ApJ, 312, 788

\noindent 
Bachiller , R. 1996, ARA\&A, 34, 111

\noindent 
Bally, J.,  Lada, C., J. 1983, ApJ, 265, 824

\noindent 
Bally, J., Devine, D., \&  Reipurth, B.
1996, ApJL, 473, L49

\noindent 
Bence, S. J., Richer, J. S., \& Padman, R. 1996, MNRAS, 279, 866

\noindent 
Dame, T. M., Ungerechts, H., Cohen, R. S., de Geus, E. J., Grennier, I. A.,

May, J., Murphy, D. C., Nyman, L.-A., \& Thaddeus, P. 1987, ApJ, 322, 706

\noindent 
Devine, D., Bally, J., Reipurth, B., \& Heathcote, S.
1997, AJ, 114, 2095

\noindent 
Eisl\"offel, J., \& Mundt, R. 1997, AJ, 114, 280

\noindent 
Gahm, G. F., Malmort, A. M. 1980, A\&A, 82, 295

\noindent 
Lada, C. J., \& Fich, M. 1996, ApJ, 459, 638

\noindent 
Fich, M. \& Lada, C. J. 1997, ApJ, 484, L63  

\noindent 
Fich, M. \& Lada, C. J. 1998, ApJS, 117, 147 

\noindent 
Lada, C, J. 1985, {\em ARA\&A}, 23, 267

\noindent 
Margulis, M., \& Lada, C. J. 1985, ApJ, 299, 925

\noindent 
Mikami, T., \& Ogura, K. 1994, MNRAS, 270, 199

\noindent 
Neckel, T., \& Klare, G. 1980, A\&AS,  42, 251

\noindent 
Ray, T.P., Eisl\"offel, J. 1994, A\&A, 290, 605

\noindent 
Reipurth, B. 1996, {\it A General Catalogue of Herbig-Haro Objects}

electronically published via anon. ftp to ftp.hq.eso.org,

directory /pub/Catalogs/Herbig-Haro

\noindent 
Reipurth, B., Bally, J., \& Devine, D.
1997, AJ, 114, 2708

\noindent 
Reipurth, B., \& Graham, J. H. 1988, A\&A, 202, 219

\noindent 
Reipurth, B., Chini, R., Kr\"ugel, E., Kreysa, E., \& Sievers, A.
1996, A\&A, 273, 221 

\noindent 
Reipurth, B., Nyman, L.Aa., \& Chini, R. 1996, A\&A 314, 258

\noindent 
Shepherd, D. S., Watson, A. M., Sargent, A. I., \&
Churchwell. E. 1998, ApJ, (in press). 

\noindent 
Stahler, S. W. 1994, ApJ, 422, 616

\noindent 
van den Bergh, S., \& Herbst, W. 1975, AJ, 80, 208

\noindent 
Zhang, Q., \& Zheng, X. 1997, ApJ, 474, 719

\newpage

\begin{tabular}[]{ l c c c c c c l }
\multicolumn{8}{l}{{\sc Table 1}:
Properties of Cloud Cores in the Circinus Cloud }\\

\hline
Designation &Location$^1$& Dimensions& $\Omega$ & M  & $\Sigma$ & $<$n($H_2$)$>$ & Comments \\
  &(\arcmin\ offsets)&(pc)&(sr)&(M$_{\odot}$)&(g~cm$^{-2}$)&(cm$^{-3}$)&  \\
\hline
\hline

Entire cloud &  -  & 36$\times$25 & 1.8$\times 10^{-3}$ & 47,000 & 
          1.1$\times 10^{-2}$ & 3.4$\times 10^1$ &  Dame et al. (1987) map \\ 
SEST map     &  -  & 7.0$\times$7.0 & 4.8$\times 10^{-5}$ & 882  & 
          7.8$\times 10^{-3}$ & 4.1$\times 10^2$ &  Region mapped at SEST\\ 
Main core    &[0,0]& 1.4$\times$3.2 & 9.1$\times 10^{-6}$ & 276  & 
          1.3$\times 10^{-2}$ & 7.0$\times 10^2$ &  Flows A, B, C \\ 
Core A       &[0,0]& 0.5$\times$0.5 & 1.5$\times 10^{-6}$ & 63   & 
          1.8$\times 10^{-2}$ & 2.1$\times 10^3$ &  Flow A only \\ 
Cir-3 core  &[19,-17]& 1.3$\times$1.3 & 3.6$\times 10^{-6}$ & 55 & 
          6.6$\times 10^{-3}$ & 5.6$\times 10^2$ &  Flow D only \\ 
SE core     &[22,-25]& 1.0$\times$1.0 & 2.0$\times 10^{-6}$ & 33 & 
          7.1$\times 10^{-3}$ & 5.4$\times 10^2$ &  Flow H \\ 

\hline
\end{tabular}

\footnotesize

Notes:

\noindent
[1]  Measured from the position of IRAS 14564--6254 (the source of flow A)
     in arc minute offsets.  

\noindent
[2]  The mass estimated from the {$^{12}$CO data of Dame et al. (1987)
     is based on the conversion factor 
     N(H$_2$) = $\rm 2.5 \times 10^{20} ~ \int T_A^*(^{12}$CO)[v] dv. 
     All other mass estimates are based on the SEST $^{13}$CO data and
     assume d~=~700~pc, N(H$_2$)/N($^{13}$CO) = $7 \times 10^5$, and
     T$_{ex}$ = 10 K.

\noindent
[3]  The densities refer to the average surface and number densities in 
     the area over which the spectral line was integrated.   For oblong
     clouds (such as the Main core) the depth of the cloud along the
     line of sight is assumed to be equal to the minor axis of the cloud.  

\newpage

\begin{tabular}[]{ l c c c c c c c c l l}
\multicolumn{10}{l}{{\sc Table 2}:
Properties of CO Outflows in the Circinus Cloud }\\

\hline
Lobe & $\Delta \alpha , ~ \Delta \delta$ $^1$ & Size & PA  & 
   v$_l$ & v$_h$ & Mass &  P &  E & \\
 & (\arcmin\ offsets) & (pc) & (deg) & 
  (km/s) &(km/s) &(M$_{\odot}$) &(M$_{\odot}$~km~s$^{-1}$) &($10^{44}$ erg) & \\
\hline
\hline

 A(b) & [0,0] & 1.3$\times$0.5& 80 &-16&-10 & 1.89 & 20.2 & 23.2& IRAS 14564--6254&, PI \\ 
      &       &               &    &   &    & 2.14 & 21.2 & 22.1& ''&, PII\\ 
 A(r) & [0,0] & 1.0$\times$0.3& 80 & -4& 3& 2.13 & 18.6 & 18.5 &  ''&, PI \\ 
      &       &               &    &   &  & 2.90 & 22.7 & 19.4 & '' &, PII\\

 B(b) & [-1,-6.5]& 1.4$\times$0.4& 177&-16&-10& 0.91& 9.7 & 11.1 & IRAS 14563--6301&, PI \\ 
      &          &               &    &   &   & 0.69& 6.3 &  6.1 & ''&, PII\\
 B(r) & [-1,-6.5]& 1.4$\times$0.4& 177& -4& 3 & 3.95 & 37.8& 39.5 &''&, PI \\ 
      &          &               &    &   &   & 3.74 & 30.0& 26.9 &''&, PII\\ 

 C(b) & [-1,-6.6]& 1.1$\times$0.3& 110 & -16& -10& 0.59 & 6.1  & 6.6 & IRAS 14563--6301&, PI \\ 
      &          &               &     &    &    & 0.50 & 4.5  & 4.2 &''&, PII\\

 C(r) & [-1,-6.6]& 0.4$\times$0.3 & 110 & -16& -10& 0.69&  8.3 & 10.0 & red lobe A&, PI\\ 
 C(r) & [-1,-6.6]& 0.4$\times$0.3 & 110 & -16& -10& 0.84& 10.2 & 12.2 & red lobe B&, PI\\ 

 C'(b) & [-1,-3.5]& 1.2$\times$0.3 & 130 & -16& -10& 0.4 &  4 & 4 & IRAS 14564--6258&, PI    \\ 
 C'(r) & [-1,-3.5]& 0.2$\times$0.5 & 130 & -4 & -3 & 0.2 &  2 & 2 & ''&, PI  \\ 

 D(b) &[19.6,-16.4]&0.5$\times$0.5 & - &-12&-8 & 0.2 & 2 & 2 & IRAS 14592--6311&, PI \\ 
 D(r) &[19.6,-16.4]&0.5$\times$0.5 & - &-4 & 0 & 0.2 & 2 & 2 & ''&, PI \\ 

 E(b) & [-0.7,4.9]&0.4$\times$0.7 & 220 & -13 & -8& 0.2 &  1.5 & 1.3 & IRAS 14563--6250&, PI    \\ 
 E(r) & [-0.7,4.9]&0.3$\times$0.4 & 220 & -5  & -2& 0.1 &  0.8 & 0.7 & ''&, PI\\

 F(b) & [3,-10]& 0.3$\times$0.3 & 100 & -13& -8& 0.1  & 0.8 & 0.7 & IRAS 14568--6304&,PI    \\ 
 F(r) & [3,-10]& 0.3$\times$0.3 & 100 & -5 & -2& 0.05 & 0.4 & 0.4 & ''&, PI \\ 

 G(b) & [-1.4,6.9]&0.4$\times$0.7 & 210 & -13 & -8& 0.2 &  1.5 & 1.3 & IRAS 14562--6248&, PI    \\ 
 G(r) & [-1.4,6.9]&0.3$\times$0.4 & 210 & -5  & -2& 0.1 &  0.8 & 0.7 & ''&, PI\\

 H(b) &[21.9,-25.1]&0.5$\times$0.5 & - &-12&-8 & 0.2 & 2 & 2 & IRAS 14596--6320&, PI \\ 
 H(r) &[21.9,-25.1]&0.5$\times$0.5 & - &-4 & 0 & 0.2 & 2 & 2 & ''&, PI \\ 

 I(b) &[11.4,-8.6] &0.5$\times$0.5 & - &-12&-8 & 0.2 & 2 & 2 & IRAS 14580--6303&, PI \\ 
 I(r) &[11.4,-8.6]&0.5$\times$0.5 & - &-4 & 0 & 0.2 & 2 & 2 & ''&, PI \\ 
\hline
\end{tabular}

\footnotesize

Notes:

\noindent
[1]  Measured from the position of IRAS 14564--6254 in arminute offsets.  

\noindent
[2] All velocities are with respect to the Local Standard of Rest (LSR).
    Momenta and kinetic energies are calculated relative to the core
    velocity which is taken to be v$_{lsr}$ = 6.5~km~s$^{-1}$.

\noindent
[3]  The average value of I($^{12}$CO)/I($^{13}$CO) in the outflow lobes
     is 20, with a minimum of 12 and a maximum of 33 over the velocity range
     covered.  Mass estimates are based on the $^{12}$CO data using the
     assumption that the $^{13}$CO intensity is I($^{12}$CO)/20.0 and 
     that the excitation conditions are as in Table 1.

\noindent
[4]  The red lobe of the C flow is confused with the B flow.  The mass
     in two sub-regions well separated from the B flow were evaluated.
     Red lobe A lies to the east of the B flow and red lobe B lies
     west of the B flow.

\noindent
[5] PI in the comments refers to Procedure I where values are calculated 
    assuming $R_{12/13}$ (=$^{12}CO$ to $^{13}CO$ ratio) is constant.  PII 
    refers to Procedure II where values are calculated assuming $R_{12/13}$
    varies with velocity. See text for further explanation.

\newpage

\begin{tabular}[]{ l c c c c c l}
\multicolumn{7}{l}{{\sc Table 3}:
Inclination Corrected Properties of the Circinus CO Outflow Lobes }\\

\hline
Lobe &  $\Delta v$ &  M  &  P  &  E  & Comments& \\
     & (km/s) &(M$_{\odot}$) &
       (M$_{\odot}$~km~s$^{-1}$) &($10^{44}$ erg)& &\\
\hline
\hline

 A  &  -18 & 0.07 & 1.35 & 2.43 & blue lobe& I \\ 
 '' &      & 0.03 & 0.53 & 0.96 & ''       & II\\
 '' &  -16 & 0.13 & 2.04 & 3.25 & ''       & I \\ 
 '' &      & 0.07 & 1.09 & 1.75 & ''       & II\\
 '' &  -14 & 0.20 & 2.85 & 3.99 & ''       & I \\ 
 '' &      & 0.14 & 1.96 & 2.75 & ''       & II\\
 '' &  -12 & 0.30 & 3.60 & 4.33 & ''       & I \\ 
 '' &      & 0.27 & 3.24 & 3.88 & ''       & II\\
 '' &  -10 & 0.46 & 4.56 & 4.56 & ''       & I \\
 '' &      & 0.65 & 6.49 & 6.49 & ''       & II\\
 '' &  - 8 & 0.73 & 5.81 & 4.65 & ''       & I \\ 
 '' &      & 0.99 & 7.88 & 6.31 & ''       & II\\

 '' & 6    & 0.89 & 5.31 & 3.19 & red lobe & I \\ 
 '' &      & 1.43 & 8.60 & 5.16 & ''       & II\\
 '' & 8    & 0.53 & 4.23 & 3.39 & ''       & I \\ 
 '' &      & 0.80 & 6.43 & 5.14 & ''       & II\\
 '' & 10   & 0.30 & 3.04 & 3.04 & ''       & I \\ 
 '' &      & 0.40 & 3.97 & 3.97 & ''       & II\\
 '' & 12   & 0.17 & 2.08 & 2.49 & ''       & I \\ 
 '' &      & 0.15 & 1.75 & 2.10 & ''       & II\\
 '' & 14   & 0.09 & 1.37 & 1.92 & ''       & I \\ 
 '' &      & 0.06 & 0.88 & 1.23 & ''       & II\\
 '' & 16   & 0.07 & 1.18 & 1.88 & ''       & I \\ 
 '' &      & 0.03 & 0.56 & 0.89 & ''       & II\\
 '' & 18   & 0.08 & 1.41 & 2.55 & ''       & I \\ 
 '' &      & 0.03 & 0.48 & 0.86 & ''       & II\\
 \hline
\end{tabular}
\vfill \eject

\begin{tabular}[]{ l c c c c c l }
\multicolumn{7}{l}{{\sc Table 3 (Continued)}} \\
\hline
Lobe &  $\Delta v$ & M  &  P  &  E  &  Comments& \\
     & (km/s)      &(M$_{\odot}$) &
       (M$_{\odot}$~km~s$^{-1}$) &($10^{44}$ erg)& \\
\hline
\hline

 B  & -18  & 0.04 & 0.75 & 1.35 & blue lobe& I \\ 
 '' &      & 0.00 & 0.00 & 0.00 & ''       & II\\
 '' & -16  & 0.06 & 0.90 & 1.44 & ''       & I \\ 
 '' &      & 0.01 & 0.22 & 0.35 & ''       & II\\
 '' & -14  & 0.09 & 1.29 & 1.81 & ''       & I \\ 
 '' &      & 0.03 & 0.45 & 0.63 & ''       & II\\
 '' & -12  & 0.13 & 1.61 & 1.94 & ''       & I \\ 
 '' &      & 0.06 & 0.76 & 0.91 & ''       & II\\
 '' & -10  & 0.21 & 2.11 & 2.11 & ''       & I \\ 
 '' &      & 0.15 & 1.52 & 1.52 & ''       & II\\
 '' & -8   & 0.38 & 3.05 & 2.44 & ''       & I \\ 
 '' &      & 0.43 & 3.40 & 2.72 & ''       & II\\

 '' &  6   & 1.30 & 7.84 & 4.68 & red lobe & I \\ 
 '' &      & 1.83 &10.97 & 6.58 & ''       & II\\
 '' &  8   & 0.87 & 6.98 & 5.59 & ''       & I \\ 
 '' &      & 0.99 & 7.89 & 6.31 & ''       & II\\
 '' &  10  & 0.58 & 5.82 & 5.82 & ''       & I \\ 
 '' &      & 0.44 & 4.39 & 4.39 & ''       & II\\
 '' &  12  & 0.42 & 4.98 & 5.98 & ''       & I \\ 
 '' &      & 0.23 & 2.74 & 3.29 & ''       & II\\
 '' &  14  & 0.31 & 4.33 & 6.08 & ''       & I \\ 
 '' &      & 0.13 & 1.79 & 2.51 & ''       & II\\
 '' &  16  & 0.25 & 3.96 & 4.37 & ''       & I \\ 
 '' &      & 0.08 & 1.27 & 2.03 & ''       & II\\
 '' &  18  & 0.22 & 3.92 & 6.98 & ''       & I \\ 
 '' &      & 0.06 & 1.00 & 1.80 & ''       & II\\
 \hline
\end{tabular}
\vfill \eject

\begin{tabular}[]{ l c c c c c l }
\multicolumn{7}{l}{{\sc Table 3 (Continued)}} \\
\hline
Lobe &  $\Delta v$ & M  &  P  &  E  &  Comments& \\
     & (km/s)      &(M$_{\odot}$) &
       (M$_{\odot}$~km~s$^{-1}$) &($10^{44}$ erg)&& \\
\hline
\hline

 C  & -18  & 0.02 & 0.29 & 0.53 & blue lobe& I \\ 
 '' & -16  & 0.03 & 0.52 & 0.83 & ''       & I \\ 
 '' & -14  & 0.04 & 0.60 & 0.84 & ''       & I \\ 
 '' & -12  & 0.08 & 0.98 & 1.18 & ''       & I \\ 
 '' & -10  & 0.13 & 1.35 & 1.35 & ''       & I \\ 
 '' & -8   & 0.29 & 2.35 & 1.88 & ''       & I \\ 

\hline 
\end{tabular}
\footnotesize

Notes:

\noindent
[1]  
$\Delta v$ is the physical velocity difference between the 
velocity bin of interest and the cloud core velocity.  It
is corrected for an assumed inclination angle of 60$^o$.  
Values are tabulated for velocity intervals of  
2 $\rm km ~s^{-1}$ 
which correspond to observed radial velocity intervals of
1 $\rm km ~s^{-1}$.

\noindent
[2]  The average value of I($^{12}$CO)/I($^{13}$CO) in the outflow lobes
     is 20, with a minimum of 12 and a maximum of 33 over the velocity 
     range covered.  Mass estimates are based on the $^{12}$CO data 
     using the assumption that the $^{13}$CO intenisty is 
     I($^{12}$CO)/20.0 and that the excitation conditions are as 
     in Table 1.

\noindent
[3] I in the comments refers to procedure I where values are calculated 
    assuming $R_{12/13}$ = I($^{12}CO$) / I($^{13}CO$) is constant.  II 
    refers to procedure II where values are calculated assuming 
    $R_{12/13}$ varies with velocity. See text for further explanation.

\newpage

\begin{tabular}[]{ l c c c c c c l }
\multicolumn{8}{l}{{\sc Table 4}:
IRAS sources in the Circinus Cloud}\\

\hline
  Source &  $\alpha$(1950) & $\delta$(1950) & 
     S(12)  & S(25) & S(60) & S(100) & Comments  \\ 
         &($^h$, $^m$, $^s$)& ($^o$, \arcmin , \arcsec ) &
   ($\mu$m) & ($\mu$m) & ($\mu$m) & ($\mu$m) &   \\ 
\hline
\hline
 
IRAS 14544--6242 & 14 54 27.3 & -62 42 53 & 0.48 & 0.34L& 2.79L& 15.77L&   \\  
IRAS 14544--6252 & 14 54 28.4 & -62 52 11 & 1.44 & 0.92 & 2.36L& 39.36L&   \\  
IRAS 14547--6302 & 14 54 44.0 & -63 02 02 & 1.73 & 0.86 & 2.25L& 33.86L&   \\ 
IRAS 14551--6248 & 14 55 10.2 & -62 48 14 & 0.40L& 0.64 & 0.88 & 38.90L&   \\   
IRAS 14556--6259 & 14 55 38.4 & -62 59 34 & 0.67 & 0.24 & 2.55L& 32.57L&   \\   
IRAS 14556--6302 & 14 55 39.9 & -63 02 38 & 1.02 & 0.81 & 1.07L& 20.46L&   \\
IRAS 14562--6248 & 14 56 12.7 & -62 48 03 & 0.25L& 1.64 & 5.26 & 94.03L& Flow G \\
IRAS 14563--6250 & 14 56 18.5 & -62 50 02 & 0.74 & 2.48 & 5.69L& 94.03L& Flow E \\
IRAS 14563--6301 & 14 56 18.6 & -63 01 42 & 1.46 & 4.08 &13.50 & 25.66 & Flows B + C \\
IRAS 14564--6258 & 14 56 25.0 & -62 58 38 & 0.41 & 1.09 &13.50L& 94.03L& Flow C'  \\ 
IRAS 14564--6254 & 14 56 28.9 & -62 54 57 & 1.38 & 7.13 &48.63 & 94.03 & Flow A \\
IRAS 14568--6304 & 14 56 51.6 & -63 04 59 & 5.04 & 9.30 &19.02 & 20.59 & Flow F, vBH65a\\
IRAS 14569--6242 & 14 56 54.3 & -62 42 35 & 1.98 & 0.50 & 2.34L& 33.13L&   \\
IRAS 14576--6312 & 14 57 36.6 & -63 12 22 & 0.28L& 0.25L& 1.37L& 14.42 &   \\
IRAS 14576--6251 & 14 57 37.0 & -62 51 11 & 1.59 & 0.71 & 2.23L& 18.93L&   \\
IRAS 14580--6303 & 14 58 04.9 & -63 03 36 & 0.28L& 1.04 & 1.68 & 20.83L& Flow I \\
IRAS 14582--6305 & 14 58 13.7 & -63 05 15 & 0.81 & 0.33L& 3.27L& 26.92L&   \\
IRAS 14583--6329 & 14 58 21.2 & -63 29 35 & 1.22 & 2.69 & 0.70 & 41.63L&   \\
IRAS 14585--6300 & 14 58 32.5 & -63 00 27 & 0.34L& 0.70L& 1.89 & 48.71L&   \\
IRAS 14591--6238 & 14 59 10.7 & -62 38 53 & 0.40 & 0.27 & 4.29L& 37.30L&   \\
IRAS 14592--6311 & 14 59 17.1 & -63 11 20 & 2.77 & 5.96 &10.54 & 11.37 & Flow D, vBH65b\\
IRAS 14593--6254 & 14 59 21.3 & -62 54 43 & 0.27L& 0.37L& 2.23 & 19.23 &   \\
IRAS 14596--6320 & 14 59 37.4 & -63 20  05& 0.32L& 0.29 & 4.02 & 16.03 & Flow H \\

\hline
\end{tabular}

\footnotesize

Notes:

\noindent
[1] Flows B and C both originate within one beam of IRAS 14563--6301.    

\newpage

\begin{tabular}[]{ l c c c c c c l }
\multicolumn{8}{l}{{\sc Table 5}:
Total Mass, Momenta, and Kinetic Energies Corrected for Hidden Mass }\\

\hline
Lobe & Size & M$_{obs}$ & M$_{tot}$ & P$_{obs}$ & P$_{tot}$ & E$_{obs}$ & E$_{tot}$ \\
     & (pc) & (M$_{\odot}$) & (M$_{\odot}$) 
     &(M$_{\odot}$~km~s$^{-1}$) &(M$_{\odot}$~km~s$^{-1}$) 
     &($10^{44}$ erg) &($10^{44}$ erg) \\
\hline
\hline

 A    & 2.1$\times$0.5  &  4.02  & 55 & 38.8 & 185 &  41.7 & 93 \\
 B    & 2.7$\times$0.4  &  4.86  & 67 & 47.5 & 227 &  50.6 & 113 \\
 C    & 1.7$\times$0.4  &  2.14  & 29 & 24.6 & 118 &  28.9 & 65 \\
 C'   & 2.1$\times$0.4  &  0.6   &  6 &  6   &  24 &   6   & 12 \\
 D    & 0.5$\times$0.5  &  0.4   &  4 &  4   &  16 &   4   &  8 \\
 E    & 0.8$\times$0.3  &  0.3   &  3 &  2.3 &   9 &   2   &  4 \\
 F    & 0.4$\times$0.3  &  0.15  & 1.5&  1.2 &   5 &   1.1 &  2 \\
 G    & 0.8$\times$0.3  &  0.3   &  3 &  2.3 &   9 &   2   &  4 \\
 H    &   ---  & $>$0.15 & $>$1.5 & $>$1.2 & $>$5 &  $>$1.1 & $>$2 \\
 I    &   ---  & $>$0.15 & $>$1.5 & $>$1.2 & $>$5 &  $>$1.1 & $>$2 \\

\hline
\end{tabular}

\footnotesize

Notes:

\noindent
[1] All quantities are corrected for an assumed inclination angle of 60$^o$
to the line-of-sight.

\noindent
[2] All quantities are summed over both the redshifted and blueshifted lobes.

\noindent
[3] The observed quantities were measured from $\Delta v_2$ = 7 km/s to 
  $\Delta v_3$ = 18 km/s.  The total quantities are extrapolated to 
  $\Delta v_1$ = 2.0 km/s. 

\newpage

\section*{FIGURE CAPTIONS}

\figcaption[Bally.fig1.ps]{
An R band ESO Schmidt photograph of the high
obscuration western portion of the Circinus cloud.
\label{fig1}}

\figcaption[Bally.fig2.ps]{
{\bf [a].}
A finder chart showing the location of CO outflows in the
western part of the Circinus complex. 
{\bf [b].}
A velocity integrated \tco\ J=1-0 map of the Circinus cloud
over the velocity range v${lsr}$ = -8 to -4 \kms\ superimposed on
the image shown in Figure 1. 
Contour levels are plotted from
1 to 14 $\rm K~km~s^{-1}$ at intervals of 
0.5 $\rm K~km~s^{-1}$.  Solid circles mark the locations
of IRAS sources.  The straight lines mark the
location and orientation of the flows discussed in the text. 
Two lines (marked C and C$'$)  indicate two sets of overlapping
outflow lobes which are hard to separate from each other.  
Therefore, for the flows labeled as C and C$'$ the tables combine
(sum) the emission from both the C and C$'$ flows.  
Squares mark the locations of 
7 Herbig-Haro objects listed in Reipurth (1994).
The axes are arc minute offsets with respect to the
position of a star located at 
$\alpha$(1950) = 14$^h$ 56$^m$ 24.9$^s$,
$\delta$(1950) = -62$^o$ 54$^{\prime}$ 59$^{\prime \prime}$
close to the millimeter wavelength continuum source.
{\bf [c].}
The \tco\ map showing the anomalous velocity
components southeast of vBH65a superimposed on the
optical image.  The velocity range of
the integration is v${lsr}$ = -10 to -8 \kms .
Contour levels are shown at intervals of 0.5 K~\kms from
0.5 to 1.5 K~\kms.
{\bf [d].}
A \co\ map of the Circinus cloud superimposed on the optical image.
Same as Figure 2a but for \co . 
Levels plotted from
0.25 to 2.5 $\rm K~km~s^{-1}$ at intervals of
0.25 $\rm K~km~s^{-1}$.  Solid circles mark the locations
of IRAS sources.  The straight lines mark the
location and orientation of the flows discussed in the text.
Two lines (marked C and C$'$)  indicate two sets of overlapping
outflow lobes which are hard to separate from each other.
Squares mark the locations of
7 Herbig-Haro objects listed in Reipurth (1994).
\label{fig2}}

\figcaption[Bally.fig3.ps]{
A \tco\ contour map of the main core that contains
the millimeter wavelength continuum source with levels plotted from
2 to 14 $\rm K~km~s ^{-1}$ at intervals of
1 $\rm K~km~s ^{-1}$.  Objects are marked as in Figure 3a.
\label{fig3}}

\figcaption[Bally.fig4.ps]{
\Tco\ maps of the Circinus cloud core.
All coordinates are arc minute offsets with respect to the
position of a star located at 
$\alpha$(1950) = 14$^h$ 56$^m$ 24.9$^s$, 
$\delta$(1950) = -62$^o$ 54$^{\prime}$ 59$^{\prime \prime}$.
 {\bf [a].} The highest velocity \Tco\ emitting gas.  Solid
 contours correspond to the blueshifted gas in the velocity 
 interval v$_{lsr}$ = -16.0 to -11.0 $\rm km~s^{-1}$. 
 Dashed lines correspond to the blueshifted gas in the velocity 
 interval v$_{lsr}$ = -1.0 to 4.0 $\rm km~s^{-1}$.
 Contour levels are shown at intervals of 1 $\rm K~km~s^{-1}$ 
 from 1 to 10 $\rm K~km~s^{-1}$.
 {\bf [b].}  
 The low velocity \Tco\ emitting gas that can just be distinguished
 from the cloud core emission.  Solid
 contours correspond to the blueshifted gas in the velocity
 interval v$_{lsr}$ = -12.0 to -10.0 $\rm km~s^{-1}$.
 Dashed lines correspond to the blueshifted gas in the velocity
 interval v$_{lsr}$ = -3.0 to -1.0 $\rm km~s^{-1}$.
 Contour levels are shown at intervals of 1 $\rm K~km~s^{-1}$ 
 from 1 to 5 $\rm K~km~s^{-1}$.
 {\bf [c].} Very low velocity \Tco\ emiting gas.  Solid
 contours correspond to the blueshifted gas in the velocity
 interval v$_{lsr}$ = -10.5 to -9.0 $\rm km~s^{-1}$.
 Dashed lines correspond to the blueshifted gas in the velocity
 interval v$_{lsr}$ = -5.0 to -3.0 $\rm km~s^{-1}$.
 Contour levels are shown at intervals of 1 $\rm K~km~s^{-1}$
 from 1 to 10 $\rm K~km~s^{-1}$.
\label{fig4}}

\figcaption[Bally.fig5.ps]{
\Tco\ spatial velocity maps of the Circinus cloud.
 {\bf [a].}  
 The A flow at PA = 80$^o$ centered on IRAS 14564--6254.
 Contour levels are shown at intervals of 0.25 K from -0.5 to 5.75 K.
 {\bf [b].}  
 The B flow at PA = 180$^o$.
 Contour levels are shown at intervals of 0.25 K from -0.5 to 5.5 K.
 {\bf [c].}  
 The C flow at PA = 120$^o$.
 Contour levels are shown at intervals of 0.25 K from -0.25 to 5.0 K.
 {\bf [d].}  
 The C$'$ flow at PA = 135$^o$.
 Contour levels are shown at intervals of 0.25 K from -0.25 to 5.0 K.
 {\bf [e].}  
 A \tco\ spatial velocity map from [25,-16] to [-10,-8] that
 illustrates the velocity structure of the core emission and
 shows the anomalous velocity component southeast of vBH65a. 
 Contour levels are shown at intervals of 0.25 K from -0.25 to 5.25 K.
\label{fig5}}

\figcaption[Bally.fig6.ps]{
Spectra of \Tco\ (thin line), \tco\ (thicker line) , and \co\ 
(thickest line) emission towards the sources of the A, B + C, 
C$'$, D, E, F, F, G, H, and I  flows obtained by forming an average 
over a 1\arcmin\ diameter region centered on each associated IRAS source.
\label{fig6}}


\figcaption[Bally.fig7.ps]{ 
Power law fits of the \Tco\ / \tco/ intensity ratio averaged over the
five outflow lobes considered in detail.  
Diamonds and crosses are used to denote the 
line wings in each region (blueshifted or redshifted).
The solid curves show the polynomial fits to the observed isotope ratio
averaged over the spatial extent of the dominant outflow lobe in each
region.   The lower right panel shows the isotope ratio for the
entire outflow complex in the core region, combining both lobes of
the A, B, and C + C$'$ outflows.   This polynomial fit to the
isotope ratio is shown as the dashed curve in the other panels.
The pair of vertical dashed lines in each panel show the velocity
extent of the emission from the Circinus cloud core.
\label{fig7}}

\figcaption[Bally.fig8.ps]{ 
Plots of the mass (M) in 2 \kms\ wide velocity intervals as 
a function of the de-projected flow velocity ($| v - v_{core} |$) 
derived using the assumption that the \Tco\ / \tco\
varies as a function of velocity as determined in Figure 7.
Symbols are as in Figure 7.  The solid lines are fits to the data
point shown.  See text for detailed discussion.
\label{fig8}}

\bigskip

\end{document}